\documentclass[aps,pra,
epsfigure,twocolumn,longbibliography,superscriptaddress]{revtex4-1}

\usepackage{amsfonts}
\usepackage{amssymb}
\usepackage{amsmath}
\usepackage{calc}
\usepackage{subfigure}
\usepackage{graphicx}
\usepackage{epstopdf}
\usepackage{dcolumn}
\usepackage{bm}
\usepackage{color} 
\usepackage{txfonts}
\usepackage[dvipsnames]{xcolor}
\usepackage{appendix}
\usepackage[colorlinks, citecolor=blue]{hyperref} 
\usepackage{soul}

\newcommand{\beq}{\begin{equation}}
\newcommand{\eeq}{\end{equation}}
\newcommand{\beqa}{\begin{eqnarray}}
\newcommand{\eeqa}{\end{eqnarray}}

\begin{document}

\title{Fast-forward scaling of atom-molecule conversion in Bose-Einstein condensates}

\author{Jing-Jun Zhu \href{https://orcid.org/0000-0002-4277-1730}{\includegraphics[scale=0.05]{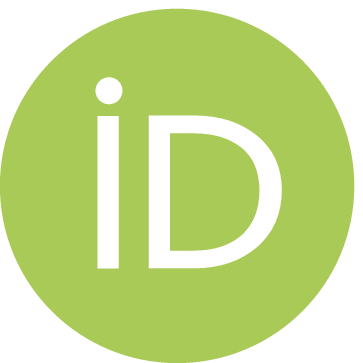}}}
\affiliation{International Center of Quantum Artificial Intelligence for Science and Technology (QuArtist) \\ and Department of Physics, Shanghai University, 200444 Shanghai, China}



\author{Xi Chen \href{https://orcid.org/0000-0003-4221-4288}{\includegraphics[scale=0.05]{ORCIDiD.eps}}}
\email{xchen@shu.edu.cn} 

\affiliation{International Center of Quantum Artificial Intelligence for Science and Technology (QuArtist) \\ and Department of Physics, Shanghai University, 200444 Shanghai, China}

\affiliation{Department of Physical Chemistry, University of the Basque Country UPV/EHU, Apartado 644, 48080 Bilbao, Spain}

\begin{abstract}

Robust stimulated Raman exact passages are requisite for controlling nonlinear quantum systems, with the wide applications ranging from ultracold molecules, non-linear optics to superchemistry. Inspired by shortcuts to adiabaticity,  we propose the fast-forward scaling of stimulated Raman adiabatic processes with the nonlinearity involved, describing the transfer from an atomic Bose-Einstein condensate to 
a molecular one by controllable external fields. The fidelity and robustness of atom-molecule conversion are shown to surpass those of conventional adiabatic passages, assisted by fast-forward driving field. Finally, our  results are extended to the fractional stimulated Raman adiabatic processes for the coherent superposition of atomic and molecular states.  
\end{abstract}

\maketitle
	\section{Introduction}
	
Over the past few decades, coherent control has
been considered {as} a strategic cross-sectional field of research for  atomic, molecular, optical physics
and photochemistry, providing {a set of quantum mechanics based methods} for the manipulation of populations 
{by laser pulses typically} \cite{Bergmann-Rev,Shapir,Shapiro-Rev,JPCreview,Guerin,STIRAPRev2,shorebook}. 
For example, the coherent control {of}  {chemical interactions} exemplifies {its} fascinating application{s} in chemistry, {for} manipulating and enchaning the product yield \cite{Brumerreview}. {Another intriguing application is} 
so-called superchemistry, in which the coherent Raman transition generates a molecular Bose-Einstein condensate (BEC) from an atomic BEC \cite{superchemstry,superchemstry2}. 
{With the advancement of modern} quantum technologies, quantum control has also emerged as an essential physical basis for state preparation 
{and manipulation} in quantum information science and quantum sensing~\cite{book,EPJD}. 

In this context, there {exist} several promising { techniques} for controlling quantum states coherently, {such as,} adiabatic passages \cite{Bergmann-Rev,Shapiro-Rev,STIRAPRev2}, composite pulses \cite{levitti,vitanovprl}, optimal control theory \cite{EPJD} and single-shot shaped pulses \cite{analytical,PRL2013D}. 
{Along with these techniques,} the concept of ``shortcuts to adiabaticity" (STA) 
provides an alternative control paradigm {that improves} the speed
and {robustness of the control process} \cite{RMP,STAreview}. 
In the case of controlled population transfer, {methods like, the} counter-diabatic (CD) driving \cite{unayan,Rice,Rice2} ({alternatively} quantum transitionless algorithm \cite{Berry,chenprl105,yichao}), invariant-based inverse engineering (IE) \cite{Xipra11,njp2012,inverse13}, fast-forward (FF) scaling \cite{masuda,kazu,nakamura} and dressed state method \cite{clerk} are capable of speeding up the conventional rapid adiabatic passage (RAP) and stimulated Raman adiabatic passage (STIRAP) 
in two- and three-level quantum system {respectively}.
Despite the experimental demonstrations in {various} quantum platforms with nitrogen-vacancy (NV) center spins \cite{Zhangprl,Zhou,Kohl}, cold atoms \cite{DuNc} 
and superconducting circuits \cite{Vepsalainen}, {it is still essential} to choose or combine these approaches, 
{depending on specific systems and objectives of the study, using appropriate features, overlaps 
and relations among them} \cite{Xipra11,Erikepra}.

Quite naturally, {the} STIRAP and its variants \cite{JPCreview,STIRAPRev2} have been {exploited to study the} magneto- and photo-association of a BEC, by using partially overlapping pulses (pump and Stokes lasers) to produce complete population transfer between two quantum states of an atom or molecule \cite{Hope,Wynarscience,drummond,Machieprl,Mackie,Lingprl,winkler}. However, the nonlinearity induced from { the} three-body collision leads to the dynamical {instability} and inefficiency with the {resulting} breakdown of adiabaticity \cite{lingpra,Jie,Itin}. In recent years, the technique of STA \cite{Dorier,Zhu},
in addition to optimal control \cite{Ronnie,Chen16} and adiabatic tracking \cite{Tracking2013,NLAdiab2016}, is
{are} considered as 
{preferable options} to {enhance} the stability and efficiency of nonlinear STIRAP. 


In this article, we explore the FF scaling of
atom-molecule conversion in BECs with {inherent} second-order nonlinearities, by extending the FF 
{to assist} {the} STIRAP \cite{masudarice,masudareview}. Using the dark state in {the} nonlinear $\Lambda$-type STIRAP as an ansatz, we construct the FF driving field {in the form of the couplings} between atomic and molecular BECs. {We prove} that the combination {of} FF field and nonlinear STIRAP overcomes the instability and inefficiency of photo- and magneto-association of atomic BEC 
by averting the unwanted diabatic transition{s}. Furthermore, the FF driving field can be similarly designed in nonlinear fractional STIRAP (f-STIRAP), generating the coherent superposition of atomic and molecular BECs. Conceptually, the FF scaling approach in nonlinear STIRAP is different from the CD driving in linear STIRAP, though they have similar 
{form} and 
{presumably similar} {physical implementation}. The instantaneous eigenstates are degenerate and non-orthogonal in the nonlinear systems, resulting in obscure calculation of CD field. {In comparison to} the original CD field, the derived FF field is also more general and efficient with extra control parameters. {Moreover, using the FF scaling approach in nonlinear STIRAP,} the state evolution always {takes place} along the dark state, {leaving the excited states unpopulated}. This provides the advantage over the IE method \cite{Dorier}, 
{where} the irreversible losses {are} inevitable. Finally, we 
{emphasize} {that} our results {could} be applicable to other nonlinear systems, {e.g.,} nonlinear optics and BEC in an accelerated optical lattice 
in the presence of the third-order Kerr-type nonlinearities.
 
In Sec. \ref{II}, we briefly review the model and Hamiltonian of nonlinear STIRAP and its variants. In Sec. \ref{III}, we derive the formula of FF driving field accordingly. In Sec. \ref{IV}, the stability and efficiency of FF assisted STIRAP and f-STIRAP are featured. Finally, we draw the conclusion in Sec. \ref{V}.

\section{Model, Hamiltonian and Adiabatic Passage}
\label{II}

\begin{figure}[t]
	\begin{center}
		\includegraphics[scale=0.65]{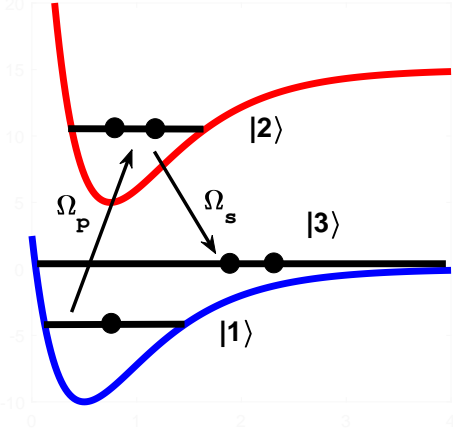}
		\caption{Schematic representation of		 coherent two-color photoassociation of  a Bose-Einstein condensate in $\Lambda$-type STIRAP, where energy level $|1\rangle$, $|2\rangle$ and $|3\rangle$ presents the electronic states for the atomic BEC, the excited and stable molecular BECs, respectively, $\Omega_{p}$ and $\Omega_{s}$ are Rabi frequencies for free-bound and bound-bound transitions.		
		}
		\label{scheme}
	\end{center}
\end{figure}

{The Schr\"{o}dinger equation,} describing the nonlinear STIRAP for coherent two-color photoassociation of a BEC in mean-field approximation as shown in Fig. \ref{scheme}, {can be expressed as following set of differential equations} \cite{Dorier}
\begin{subequations}
	\label{Horginal}
	\begin{align}
	&i \dot{c}_{1}=K_{1}c_{1}+\Omega_{p}\bar{c}_{1}c_{2},\\
    &i\dot{c}_{2}=K_{2}c_{2}+\Delta_{p}c_{2}+\Omega_{p}c_{1}^{2}+\Omega_{s}c_{3},\\
	&i\dot{c}_{3}=K_{3}c_{3}+(\Delta_{p}-\Delta_{s})c_{3}+\Omega_{s}c_{2},
	\end{align}
\end{subequations}
where $\Omega_{p,s} \equiv \Omega_{p,s}(t)$ are the
time-dependent Rabi frequencies of pump and Stokes fields for free-bound and bound-bound transitions { respectively,} $\Delta_{p,s}$ represent the { corresponding} detunings, and {$c_{j}$} are the amplitude in state $|j\rangle$. Here the overdot represents the derivative with respect to time. Typically, the process of photoassociation {aims} to 
remove two atoms from state $|1\rangle$, 
and create a stable molecule in state $|3\rangle$,
by using the two lasers coupling with a bound-bound  molecule in excited state $|2\rangle$. The most intriguing property presented here comes from the second-order nonlinearity {which appears in the form of} pump coupling {as well as the} third-order Kerr-type nonlinearity,
$K_{i}=\sum_{j=1}^{3}\Lambda_{ij}|c_{j}|^2$, with $\Lambda_{ij}$ {being} some {system dependent} constants and populations $|c_{j}|^2$. 
In the context of atom-molecular conversion in BECs, the extra $c_{1}$ and $\bar{c}_{1}$ terms 
{appearing in front of}  $\Omega_{p}$ describe the {$1:2$} resonance between the {ground} atomic state and the excited molecular states. 
The total population is conserved and is normalized as $|c_{1}|^{2}+2(|c_{2}|^{2}+|c_{3}|^{2})=1$. Here 
we write down the Hamiltonian (\ref{Horginal}) after performing a change of variable $c_{2,3}\mapsto c_{2,3}/\sqrt{2}$, yielding the usual normalization, $|c_{1}|^{2}+|c_{2}|^{2}+|c_{3}|^{2}=1$ for convenience.

Since the resonance-locking condition, $\Delta_{p}=2K_1-K_2$ and $\Delta_s= K_3-K_2$, compensates Kerr nonlinear terms with detunings \cite{Dorier},  we simplify the previous Hamiltonian,
see Eq. (\ref{Horginal}), within the on-resonance condition 
as 
\begin{subequations}
	\label{Hresonance}
	\begin{align}
	&i\dot{c}_{1}=\Omega_{p}\bar{c}_{1}c_{2},\\
	&i\dot{c}_{2}=\Omega_{p}c_{1}^{2}+\Omega_{s}c_{3},\\
	&i\dot{c}_{3}=\Omega_{s}c_{2},
	\end{align}
\end{subequations}
where the second-order nonlinearities are still involved 
which may lead to the dynamical instability \cite{Dorier,Zhu}. In principle, 
there may exist more non-orthorgonal eigenstates 
than the dimension of the
Hilbert space in nonlinear systems \cite{Jie}. 
Nevertheless, in analogy to {its} linear counterpart \cite{Bergmann-Rev,STIRAPRev2}, the nonlinear $\Lambda$-type STIRAP still supports a dark state (or so-called the coherent population
trapping state) with zero eigenvalue \cite{Machieprl,Mackie} 
which is decoupled from the excited state. Therefore, by setting $c^0_2 \simeq 0$, and using $|c^0_1|^2 + |c^0_3|^2=1$, we obtain the instantaneous population, 
\begin{equation}
|c^0_1|^2=1- |c^0_3|^2=  \frac{2\Omega_{s}}{\Omega_s+\Omega_e},
\end{equation} 
from which the dark state, corresponding to the eigenvector $|\Psi_0(t)\rangle=[c^0_1,c^0_2,c^0_3]^{T}$, 
is calculated as 
$|\Psi_0(t)\rangle= \mathcal{N} (\Omega_s |1\rangle - c^0_1 \Omega_p|3\rangle)$, 
with $\Omega_e ={(\Omega_{s}^{2}}+4\Omega_{p}^{2})^{1/2}$ and {$ \mathcal{N}$ being} the normalization constant. As used in conventional linear STIRAP \cite{Bergmann-Rev,STIRAPRev2},
the dark state is further reformulated into
\beq
\label{darkstate}
|\Psi_0(t)\rangle = \cos \Theta (t) |1\rangle - \sin \Theta(t) |3\rangle,
\eeq
with the mixing angle,
\beq
\label{mixing}
\Theta(t) =\arctan\left(\frac{c^0_1 \Omega_p}{\Omega_s}\right)= \frac{\sqrt{2}\Omega_p}{\sqrt{\Omega_s(\Omega_s+\Omega_e)}}.
\eeq  
This dark state has {already} been experimentally verified, {e.g.}, through the superposition state of atomic  and molecular BECs  \cite{winkler}.  
Apart from it, we have the other two eigenstates: $[0, \pm 1/\sqrt{2}, 1/\sqrt{2}]^{T}$, with the eigenvalues being $\pm \Omega_s/2$. When $\Omega_s/\Omega_p<1/\sqrt{2}$, two more eigenstates exist
$[(1/2-\Omega^2_{s}/\Omega^2_{p})^{1/2},\pm 1/\sqrt{2}, \Omega_{s}/\Omega_{p}]^{T} $ {which has the eigenvalues}  $\pm \Omega_p/ \sqrt{2}$. 
Due to the lack of orthorgonality between the dark state and the other eigenstates, the usual adiabatic condition {for} linear STIRAP {does not hold in the nonlinear case}. Thus, one can apply the 
linear stability analysis {only} around the fixed stable point \cite{Lingprl}, {which corresponds} to the dark state, {for calculating} three orthogonal eigenstates, 
$w_0= \mathcal{N}_0 [-\Omega_{s}/2, 0, c^0_1\Omega_{p} ]^{T}$, and $w_{\pm} =  \mathcal{N}_{\pm} [c^0_1\Omega_{p}, \epsilon_{\pm}, \Omega_{s}]^{T} $. {The eigenvalues corresponding to $w_{0,\pm}$ are} $\epsilon_{0}=0$ and
$\epsilon_{\pm}=\pm \sqrt{\Omega_{s}\Omega_{e}} $ 
respectively. {$\mathcal{N}_{0,\pm}$ are the normalization constants}. {Accordingly}, the adiabatic condition, suitable for the nonlinear STIRAP, is derived as \cite{Lingprl},
\beq
\label{adcondition} 
A_{nl} \approx \left(\frac{\mathcal{N}_{+}}{w_{+}}+ \frac{\mathcal{N}_{-}}{w_{-}}\right)^{1/2} \left(\frac{\dot{\Omega}_{p}\Omega_{s}-\dot{\Omega}_{s}\Omega_{p}}{\Omega_{s}+\Omega_{e}}\right)\ll 1.
\eeq
{It is straightforward to observe from the} adiabatic approximation {in Eq. (\ref{adcondition}),} the pump and Stokes Gaussian pulses \cite{Bergmann-Rev,STIRAPRev2}, {chosen as}
\begin{subequations}
\label{Gaussian0}
\begin{align}
&\Omega_{p}(t)=\Omega_{0}  e^{-(t-\tau)^{2}/T^{2}},\\
&\Omega_{s}(t)=\Omega_{0}e^{-(t+\tau)^{2}/T^2},
\end{align}
\end{subequations}
transfer {the population} from state $|1\rangle$ at initial time $t=t_i$ to $|3\rangle$ at final time  $t=t_f$ along the dark state (\ref{darkstate}). 
{Note} that $\tau$, $T$, $\Omega_{0}$ {represent} the
center, width and amplitude of the Gaussian pulses { respectively}. 
{A more general case would be a combination of} one pump and two Stokes Gaussian pulses \cite{fractional},
\begin{subequations}
	\label{Gaussian-f}
	\begin{align}
		&\Omega_{p}(t)=\Omega_{0} \sin \beta e^{-(t-\tau)^{2}/T^{2}},\\
	&\Omega_{s}(t)=\Omega_{0}e^{-(t+\tau)^{2}/T^2}+\Omega_{0} \cos \beta e^{-(t-\tau)^{2}/T^{2}},
	\end{align}
\end{subequations}
which asymptotically {becomes,}
\beq
\label{asymp}
0 \stackrel{t=-\infty}{\longleftarrow} \frac{\Omega_p(t)}{\Omega_s(t)} \stackrel{t=+\infty}{\longrightarrow} \tan \beta.
\eeq
{$\Omega_p(t)$ and $\Omega_s(t)$ creates} the coherent superposition of $|1\rangle$ and $|3\rangle$ adiabatically by using {the} dark state. The constant, $\beta$, can be determined by the final target state 
{through} {the combination of} Eqs. (\ref{mixing}) and (\ref{asymp}).
{Specially} for the nonlinear f-STIRAP \cite{Mexico_N}, one has to set $\beta=\arctan{\sqrt{2}}$ 
when the conditions $\Theta(t_f) = \pi/4$ and $c^0_1(t_f)=1/\sqrt{2}$ are stipulated to guarantee the state {to be a} superposition of $|1\rangle$ and $|3\rangle$ with equal amplitude.  {Generally it takes long time to evolve the system when the adiabatic condition (\ref{adcondition}) is satisfied,} {thus spoiling the state by decoherent effect or repeating operation with more energy cost. {In what follows, we shall develop the FF scaling approach to speed up the nonlinear STIRAP and f-STIRAP, thus circumventing such difficulties.}

\section{Fast-Forward Scaling Approach}
\label{III}

In this section, we {shall} generalize the FF scaling approach, with the motivation to accelerate the nonlinear STIRAP or {f-STIRAP,} subjected to a slow variation of pulses. Inspired by {the fundamental work of Masuda and Nakamura} \cite{masuda}, 
the FF field for accelerating adiabatic processes { have} been carried out for {several examples like,} 
discrete  multi-level quantum system \cite{masudarice,masudareview} and the nonlinear Gross-Pitaevskii {equation} or the corresponding Schr\"{o}dinger equation \cite{Masudapra2008,Masudapra2011}.
In order to {recapitulate} the FF scaling for our proposal, we choose the dark state (\ref{darkstate}) as an Ans\"{a}tz, 
\beq 
\label{ff}
|\Psi_{FF}(t)\rangle = \cos[\Theta (R(t))] e^{i f_1 (t)}|1\rangle - \sin[\Theta(R(t))] e^{i f_3 (t)} |3\rangle,
\eeq
with {$R(t)$ being the ``magnification
factor" for the rescaled time}. The phase factors $f_{1,3}(t)$ {are introduced to satisfy} the  time-dependent Schr\"{o}dinger equation,
$i \partial_t |\Psi_{FF} (t)\rangle = H_{FF}(t) |\Psi_{FF}(t)\rangle$, {which comes out to be,}
\begin{subequations}
	\label{H}
	\begin{align}
	&i\dot{c}_{1}=\Omega^{FF}_{p}\bar{c}_{1} c_{2}+\Omega^{FF}_{c}c_{3},\\
	&i\dot{c}_{2}=\Omega^{FF}_{p}c_{1}^{2}+\Omega^{FF}_{s}c_{3},\\
	&i\dot{c}_{3}=\Omega^{\ast FF}_{c}c_{1}+\Omega^{FF}_{s}c_{2},
	\end{align}
\end{subequations}
with the modified Rabi frequencies of pump, Stokes and an additional FF fields being  
 $\Omega^{FF}_{p} \equiv \Omega^{FF}_p(t)$, $\Omega^{FF}_{s} \equiv \Omega^{FF}_s(t)$, $\Omega^{FF}_c \equiv\Omega^{FF}_c(t) $ { respectively}. 
Here the {boundary} conditions $f_{1,3}(t_i)=f_{1,3}(t_f)=0$ are {required to connect to the corresponding} adiabatic reference.
By inserting this ansatz (\ref{ff}) into the dynamical equation, we have the following equations:
\begin{eqnarray}
\frac{\Omega^{FF}_{p}(t)}{\Omega^{FF}_{s}(t)} &=&\frac{\sin[\Theta(R(t))]}{\cos^{2}[\Theta(R(t))]}e^{i [\Delta f(t)-f_1(t)]},
\\
\frac{d\Delta f(t)}{dt} &=&\left\{\frac{2 \cos[ 2\Theta(R(t))]}{\sin [2\Theta(R(t))]}\right\}\mathrm{Re}[\Omega^{FF}_{c}e^{i\Delta f(t)}],
\\
\frac{\partial \Theta}{\partial R} \frac{\partial R}{\partial t} &=& \mathrm{Im}[ \Omega^{FF}_{c}e^{i\Delta f(t)}].
\end{eqnarray} 
The Rabi frequencies in the FF scaling approach are finally obtained as 
\begin{eqnarray}
\label{OmegaFF}
\frac{\Omega^{FF}_{p}(t)}{\Omega^{FF}_{s}(t)} &=&\frac{\Omega_{p}(R(t))}{\Omega_{s}(R(t))}e^{i[\Delta f(t)-f_1(t)]},
\\
\label{FFfield}
\Omega_{c}^{FF}(t)&=&e^{-i\Delta f(t)}\left\{\frac{\sin[2\Theta(R(t))]}{2\cos[2\Theta(R(t))]}\frac{d\Delta f(t)}{dt}+i\frac{\partial \Theta}{\partial R} \frac{\partial R}{ \partial t}\right\},~~~
\end{eqnarray}
with $\Delta f(t)=f_{3}(t)-f_{1}(t)$. When $f_{1,3}(t)=0$ is further assumed, the Rabi frequencies can thus {be} simplified as 
\begin{eqnarray}
\label{FF-1}
\frac{\Omega^{FF}_{p}(t)}{\Omega^{FF}_{s}(t)} &=&\frac{\Omega_{p}(R(t))}{\Omega_{s}(R(t))},
\\
\label{FF-2}
\Omega_{c}^{FF}(t) &=& i\frac{\partial \Theta}{\partial R} \frac{\partial R}{ \partial t}.
\end{eqnarray}
Obviously, the additional FF driving field $\Omega_{c}^{FF}(t)$ {is dependent} on the magnification factor {and essential} for the acceleration of adiabatic passages. 
The pump and Stokes fields, {after FF scaling, constitute} the same 
{ratio} but with the rescaled time.
{For} $R(t) = \eta t$, 
{when} the rate of change in $R(t)$ {is small}, { i.e.,} $\eta \ll 1$, the adiabatic process is recovered 
The FF field {vanishes in this limit}, i.e. $\Omega_{c}^{FF}(t) \simeq 0$. In the case of $R(t)=t$, the FF driving field can be written as $\Omega_{c}^{FF}(t)=i \dot{\Theta}$, 
which is {similar to the} CD driving in linear STIRAP \cite{chenprl105}. However, the mixing angle {is associated with the $c^0_1$} in Eq. (\ref{mixing}) { which} results in different auxiliary interaction  between  $|1\rangle$ and $|2\rangle$, {distinguishing it from its} linear counterpart.

\begin{figure}[t]
	\begin{center}
		\includegraphics[scale=0.45]{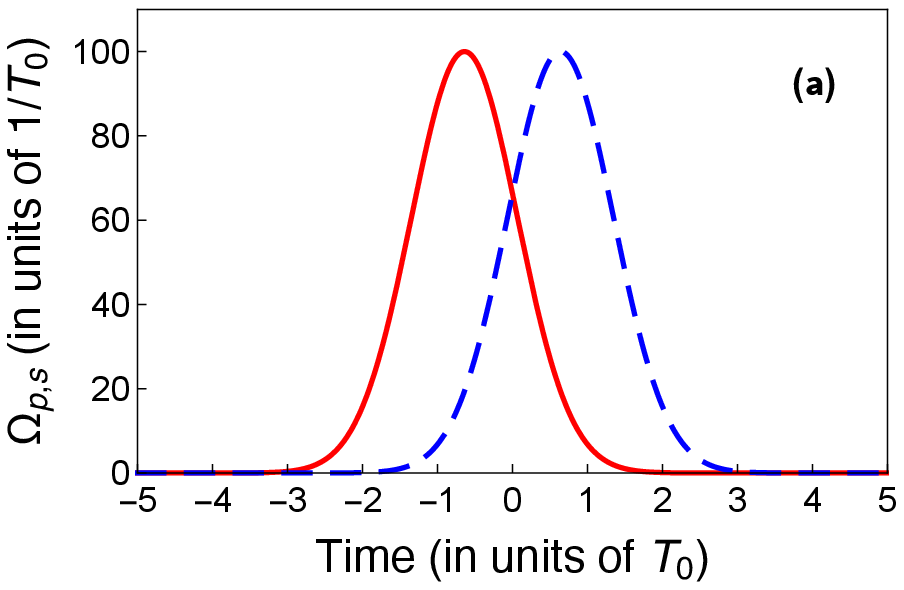}
		\includegraphics[scale=0.45]{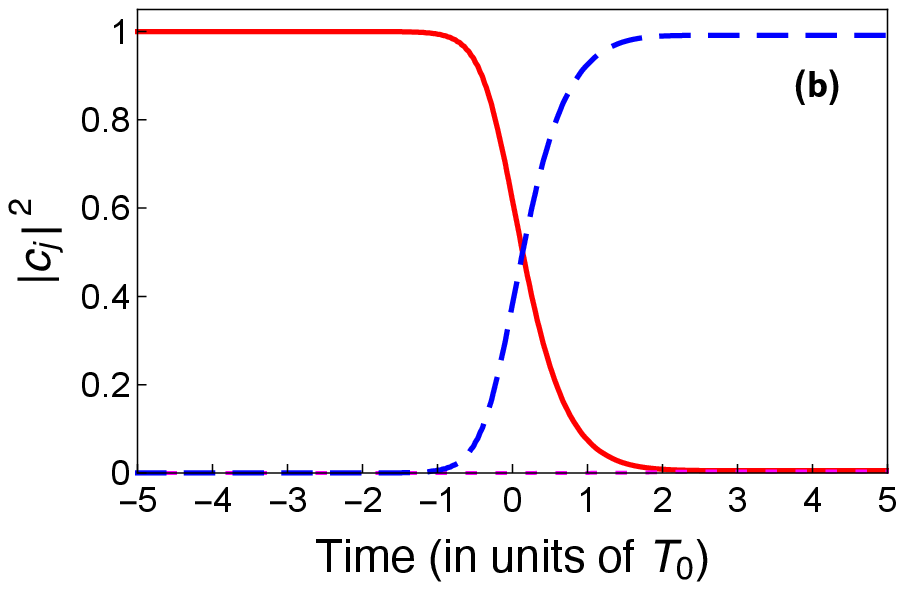}\\
		\includegraphics[scale=0.45]{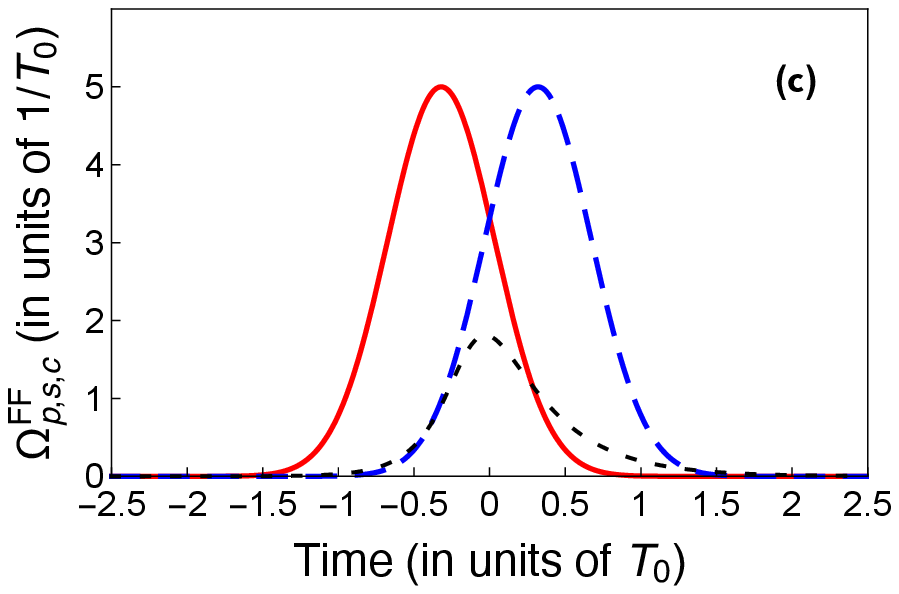}
		\includegraphics[scale=0.45]{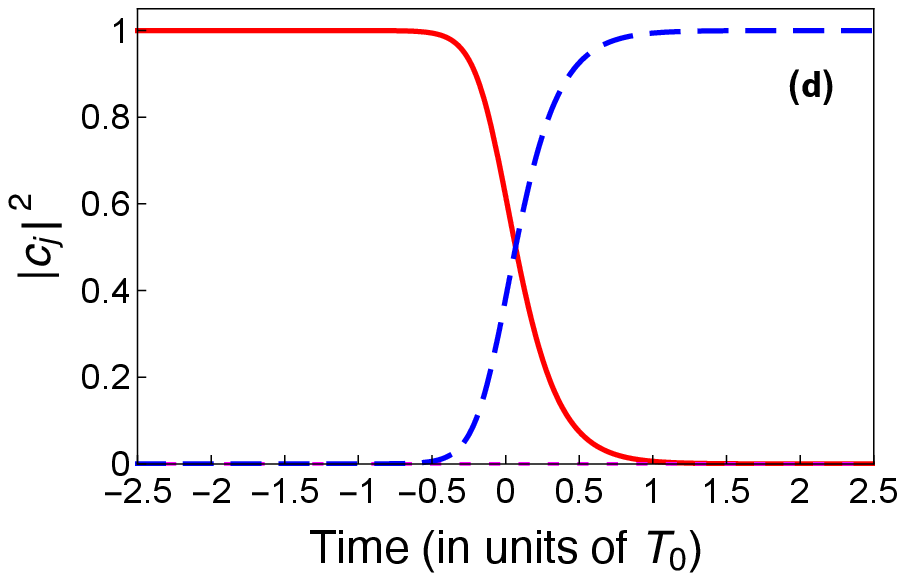}
		\caption{Nonlinear STIRAP (a) for state transfer (b) from an atomic BEC to a molecular BEC  is presented, where $\Omega_{s}$ (solid red) and  $\Omega_{p}$ (dashed blue) sequences
			are Gaussian type with the amplitude $\Omega_{0}=100$ (in the units of $1/T_0$), $\tau= 0.64 T$, $T=1$ 
			(in the units of $T_0$) and $T_0=3.1 \times 10^{-5}$s. The final population $|c_3|^{2} = 0.9914$ is achieved with total time $10T$. For comparison,
			FF-assisted STIRAP (c) for state transfer (d) is also presented, where $\Omega^{FF}_{s}$ (solid red) and  $\Omega^{FF}_{p}$ (dashed blue) with $\Omega_{0}=5$ and other parameters are the same. 
			Assisted by the FF driving field $\Omega^{FF}_c$ (dotted black), the final population $|c_3|^{2} = 0.9999$ is achieved with total time $5T$.} 
		\label{fig2}
	\end{center}
\end{figure}

\begin{figure}[t]
	\begin{center}
		\includegraphics[scale=0.65]{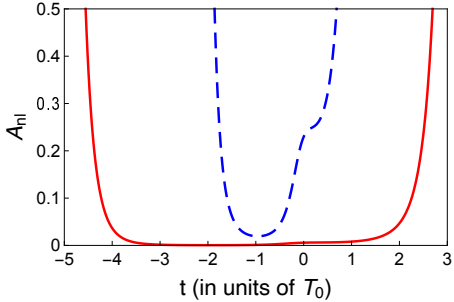}
		\caption{Parameter $A_{nl}$ is depicted to quantify the adiabatic condition (\ref{adcondition}), where the parameters are used for nonlinear STIRAP (solid red) and FF-assisted STIRAP (dashed blue) in Fig. \ref{fig2}. 	
		}
		\label{fig3}
	\end{center}
\end{figure} 

{It is important to note that, even though the similarities are predominant} {between the FF scaling, presented here, and other traditional STA methods like CD driving and IE methods, the differences between them are also significant.} Though the additional coupling between $|1\rangle$ and $|3 \rangle$ are required for both
FF scaling and CD driving methods, the FF scaling approach is more general in the sense that when
$f_{13}(t) \neq 0$, the FF field has both real and imaginary parts {with} pulse area larger than $\pi$ \cite{masudarice}. {Most} importantly, FF scaling approach is {fundamentally} different from CD driving. In order to obtain the CD term, one has to calculate the { non-adiabatic contribution} after diagonalizing {the} Hamiltonian, {which is rather straighforward to calculate as the eigen-spectrum is known in linear STIRAP}. {On the contrary,} in nonlinear STIRAP, we can't use {the} eigenstates and their orthogonality to obtain the CD driving directly. Instead, we assume the dark state as an ansatz for constructing the FF driving field, such that the state transfer is always along the dark state. This also provides {significant} advantage over the IE method used in Ref. \cite{Dorier}, in which the intermediate state $|2\rangle$ is populated, thus leading to the inevitable losses. Moreover, various functions of $R(t)$ can be further adopted for accelerating the adiabatic passage, which provides more flexibility as well \cite{Masudapra2008,Masudapra2011}. By selecting the magnification factor $R(t)$ for the rescaled time, the FF scaling is closely connected with the time-rescaled method, recently proposed in \cite{prr_RT}. However, we should point out that the time-rescaled dynamics, driven by the modified fields (\ref{FF-1}), works perfectly only when the original protocol is adiabatic,
since the adiabatic condition cannot be improved 
without the auxiliary coupling, see Eq. (\ref{FF-2}).


\section{Efficiency and Stability}
 
 \label{IV}
 

We first check the conventional nonlinear STIRAP in {aforementioned} $\Lambda$-type nonlinear system. By using Gaussian shapes of pump and Stokes fields, see Eq. (\ref{Gaussian0}), the state can be adiabatically transferred from an atomic BEC at initial time $t_i=-5T$ to a molecular one at $t_f=5T$ , as shown in Fig. \ref{fig2} (a,b), where $\Omega_0 =100$ (in the units of $1/T_0$), $\tau=0.64 T$, $T=1$ (in the units of $T_0$). Normally, $T_0=3.1 \times 10^{-5}$ s can be chosen in the practical experiment \cite{Wynarscience}, corresponding to $\Omega_{0}= 3.226$ MHz. The final population $|c_3(t_f)|^2 = 0.9914$ is achieved without exciting {the} state $|2\rangle$ when the total time is $10T$ with the fixed  $\Omega_0 =100$. Remarkably, the modified pump/Stokes fields and FF driving field are designed to accelerate the nonlinear STIRAP, as shown in Fig. \ref{fig2} (c,d), where 
 $\Omega_{0}=5$, and other parameters are the same as those in Fig. \ref{fig2} (a,b). By introducing
 $R(t)=at$, the total {evolution} time is decreased up to $5T$ with $a=2$,
 when assisted by the FF field. From the comparison in Fig. \ref{fig2}, we demonstrate that
 the assisted FF driving field really speed up the original nonlinear STIRAP, {by} following the dark state, as seen in population evolution. The final population reaches $|c_3|^2=0.9999$ in Fig. \ref{fig2} (d), {even} when the parameters {do not} fulfill the adiabatic condition (\ref{adcondition}). 
In order to quantify the acceleration, we calculate the adiabatic condition (\ref{adcondition}), see Fig. \ref{fig3}, where $A_{nl} \ll 1$, {for} the parameters {that are} used in conventional nonlinear STIRAP.
{When the transfer time is shortened, corresponding parameters make $A_{nl}$ significantly large so that the adiabaticity is broken,} with lower intensities of the pulses. {For simplicity,} one can {choose} $R(t)=t$  
while keeping the total time {at} $10 T$. 
{It can still be shown} that the auxiliary FF field assists {the pump and Stokes fields} to achieve the high-fidelity state transfer with $\Omega_{0}=5$. Here, {the} Gaussian pulses {do not} fulfill the adiabatic condition {and} the FF driving field speeds up {the} nonlinear STIRAP when $R(t)=t$, {reducing the system evolution time for} small $\Omega_{0}$. This also clarifies the importance of auxiliary coupling, and makes the difference from time-rescaled method \cite{prr_RT}, as mentioned before.
	
\begin{figure}[t]
	\begin{center}
		\includegraphics[scale=0.45]{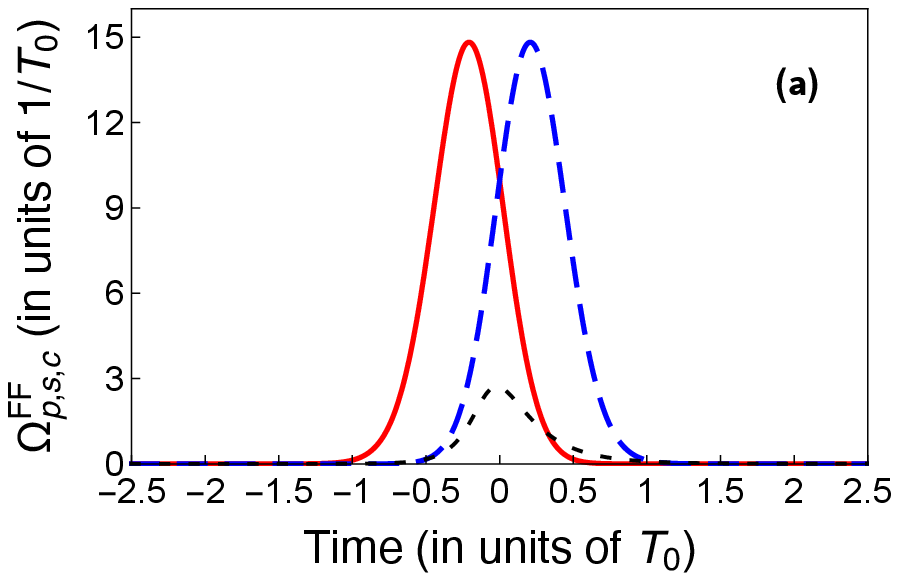}
		\includegraphics[scale=0.45]{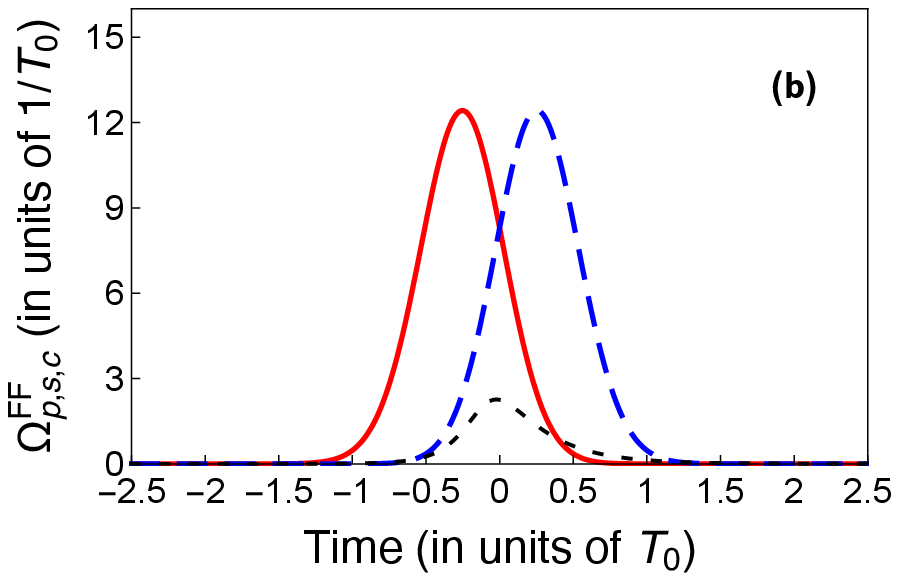}
		\caption{By using different $R(t)$, the modified pump and Stokes Gaussian pulses are presented for nonlinear STIRAP (a,b), together with the assisted FF driving field, where $\Omega^{FF}_{s}$ (solid red)  $\Omega^{FF}_{p}$ (dashed blue) and $\Omega^{FF}_{c}$ (dotted black). 
			The parameters are the same as those in Figs. \ref{fig2}. Here (a) and (b) correspond to the trigonometric 
			and polynomial Ans\"{a}tzes of $R(t)$ respectively.}
		\label{fig4}
	\end{center}
\end{figure}

Moreover, we select other functions of $R(t)$, demonstrating the diversity. For instance, one option is trigonometric Ans\"{a}tz \cite{prr_RT}, 
\beq
\label{Rsin}
R(t)=at-\frac{t_{f}-t_{i}}{2\pi a}(a-1)\sin \left[\frac{2\pi a}{t_{f}-t_{i}} \left(t-\frac{t_{i}}{a}\right)\right],
\eeq
where its inverse function and first derivative 
satisfy the following boundary conditions, $R^{-1}(t_{i})=t_{i}/a $, $R^{-1}(t_{f})=t_{f}/a$, and $R'(t_{i})=R'({t_{f}})=1$. Here the additional conditions on first derivative imply that the time-rescaled Hamiltonian coincides with the original one at initial and final times. Fig. \ref{fig4} (a) shows
the modified pump, Stokes Gaussian pulses for nonlinear STIRAP, and the assisted FF driving field. Here we use $a=2$ to compare the results obtained from $R(t)=2t$, by keeping the same fidelity at $t=t_f$ as that in Fig. \ref{fig2}.
We find that the amplitude of Gaussian pulses becomes larger by using the trigonometric function,
while the dynamics of population (not shown here) does not change too much.  Alternatively, the forth degree polynomial,
$ R(t) = \sum^{3}_{i=0} \eta_j t^{i} $, can be adopted as well, where the four coefficients $\eta_j$ are completely solvable by using the aforementioned boundary conditions. We do not write them down explicitly, to avoid the lengthy expression. 
In this case, the modified pump and Stokes Gaussian pulses are presented in Fig. \ref{fig4} (b), together with the corresponding FF driving field. It is confirmed by the comparison that the proper choice of $R(t)$  help decrease the amplitude of pulses. 

 \begin{figure}[t]
	\begin{center}
		\includegraphics[scale=0.45]{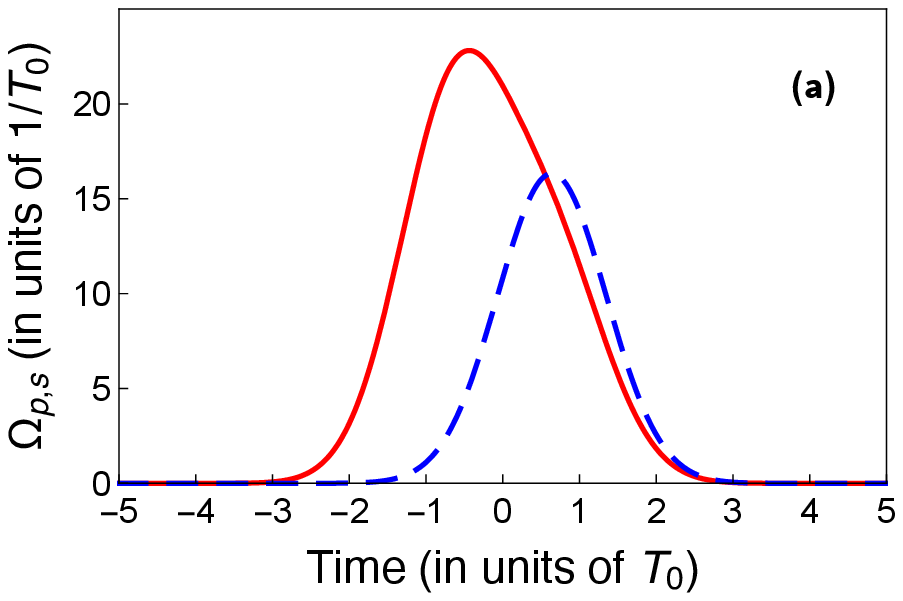}
		\includegraphics[scale=0.45]{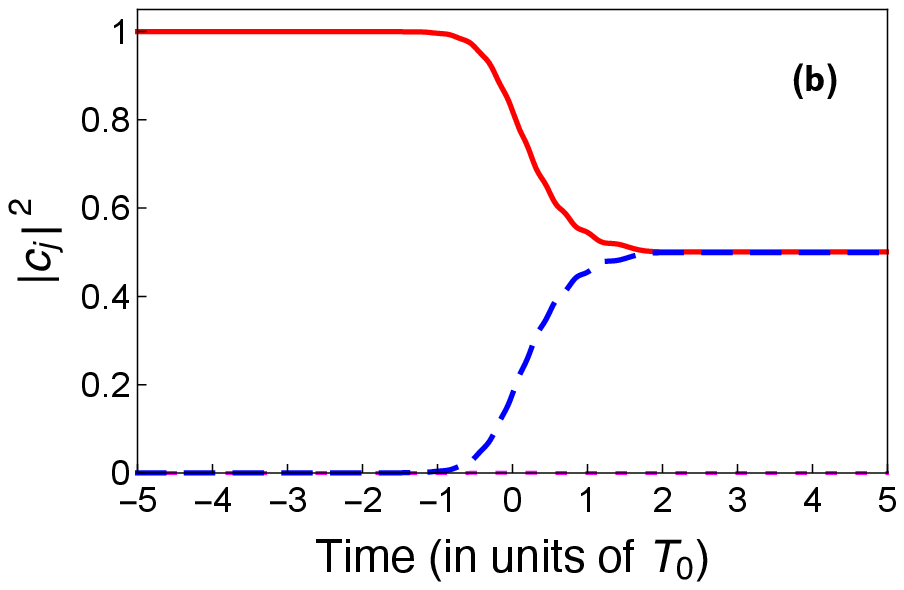}
		\\
		\includegraphics[scale=0.45]{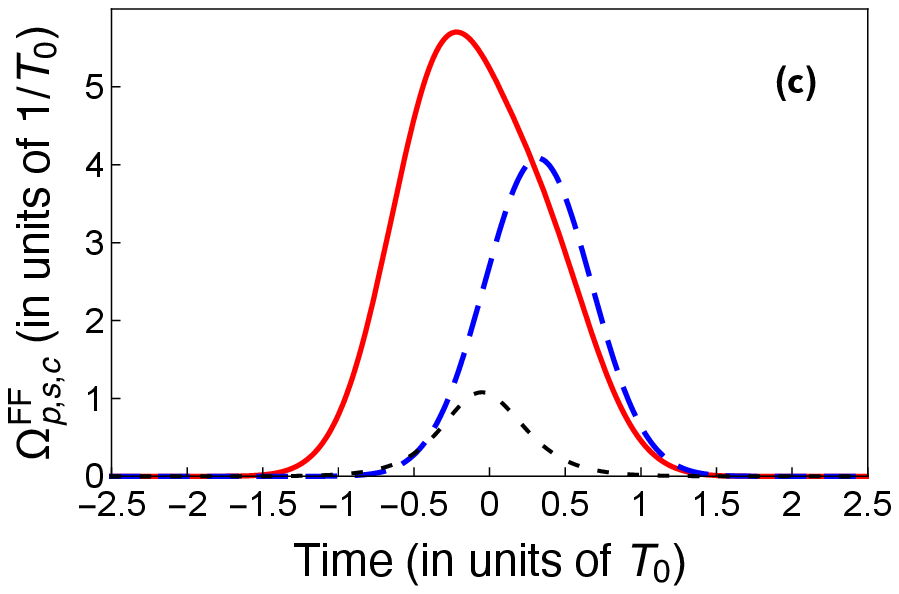}
		\includegraphics[scale=0.45]{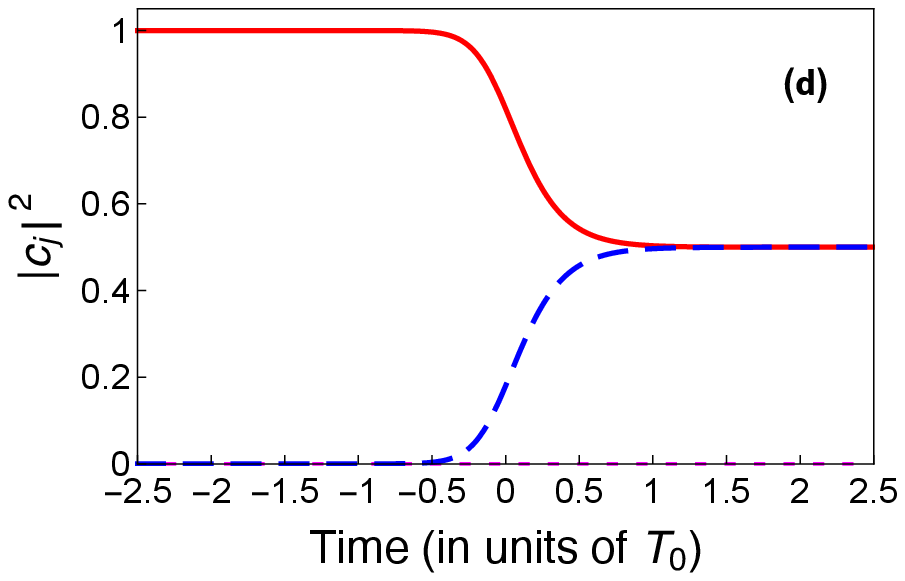}
		\caption{Nonlinear f-STIRAP (a) for generating the coherent state superposition (b) is presented, where $\Omega_{s}$ (solid red) and  $\Omega_{p}$ (dashed blue) sequences are Gaussian type with the amplitude $\Omega_{0}=20$ (in the units of $1/T_0$), $\tau= 0.64 T$, $T=1$ (in the unites of $T_0$) and $T_0=3.1 \times 10^{-5}$s. The superposition of an atomic BEC and a molecular BEC with equal amplitudes is provided with total time $10T$. For comparison, FF-assisted f-STIRAP (c) for generating such superposition (d) is also presented, where $\Omega^{FF}_{s}$ (solid red) and  $\Omega^{FF}_{p}$ (dashed blue) with 
			{$\Omega_{0}=5$} and other parameters are the same. Assisted by the FF driving field $\Omega^{FF}_c$ (dotted black), the superposition of atomic and molecular BECs with equal amplitudes is achieved with total time $5T$.} 
		\label{fig5}
	\end{center}
\end{figure}

Next, we also apply the  FF driving  field for accelerating nonlinear f-STIRAP. In Fig. \ref{fig5} (a,b) we recover the original adiabatic process to generate the coherent superposition of an atomic BEC and a molecular BEC with equal amplitude 
where Gaussian pulses of pump and Stokes fields are used, see Eq. (\ref{Gaussian-f}), with $\Omega_0=20$ and $\beta=\sqrt{2}$. Similar to {the} nonlinear STIRAP, we apply the FF driving field {along with the} modified pump and Stokes fields to speed up the nonlinear f-STIRAP, see in Fig. \ref{fig5} (c,d). Here we choose the rescaling function as $R(t)=2t$ for simplicity, 
which can shorten the total evolution time from $10T$ to $5T$, with small {coupling} amplitude $\Omega_{0}=5$. With the assisted FF driving field, the state evolution follows exactly the adiabatic reference in Fig. \ref{fig5} (b,d) to achieve the perfect coherent superposition with the ratio $1:1$, but without populating excited state $|2\rangle$. {It is evident from the} Fig. \ref{fig5} that the FF scaling approach {provides desired state superposition,} { which can be generalized for} other ratio of amplitudes {as well} by changing the parameter $\beta$ in Eq. (\ref{Gaussian-f}) through the mixing angle (\ref{mixing}). 

\begin{figure}[t]
	\begin{center}
		\includegraphics[scale=0.45]{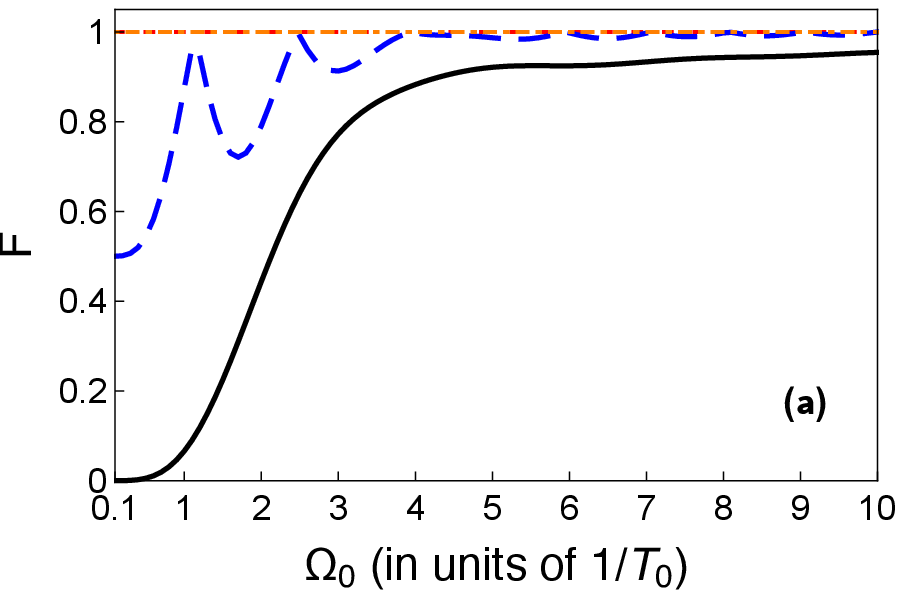}
		\includegraphics[scale=0.45]{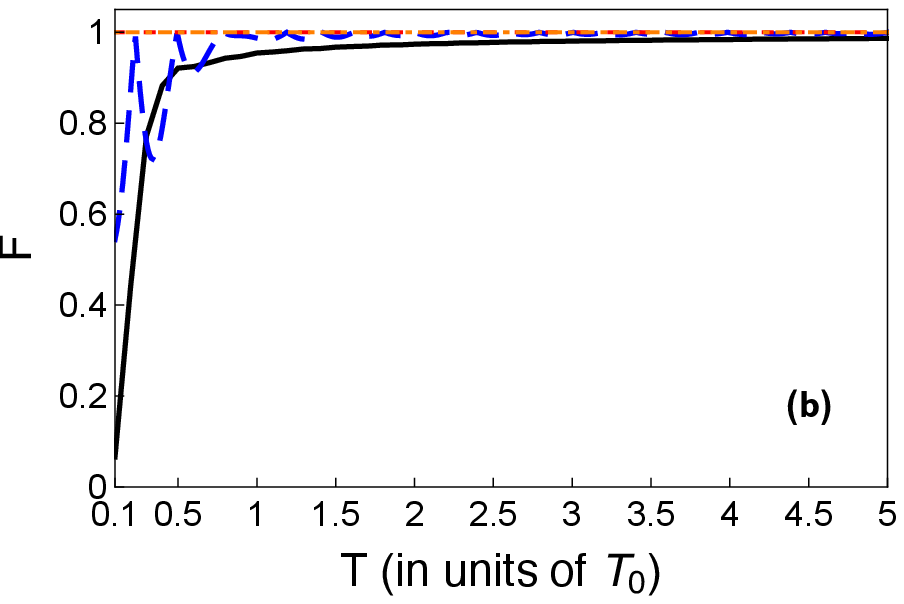}
		\caption{Fidelity, $F= |\langle \Psi_0 (t_f) | \Psi (t_f) \rangle|^2$, versus the amplitude $\Omega_{0}$ (a) and the width $T$ (b) of Gaussian pulses, where $\Psi_0 (t_f) $ is target state and $ |\Psi (t_f) \rangle$ is the final solution of time-dependent Schr\"{o}dinger equation. Here the fidelities of nonlinear STIRAP, nonlinear f-STIRAP and FF-assisted STIRAP are denoted by solid black, dashed blue and dotted red lines, and that of FF-assisted f-STIRAP (dash-dotted orange) are undistinguishable. Other parameters are the same as those in Figs. \ref{fig2} and \ref{fig5}.}
		\label{fig6}
	\end{center}
\end{figure} 

Finally, we address the issue on efficiency and stability of nonlinear STIRAP and f-STIRAP assisted by the FF driving field.  
{Fig. \ref{fig6} (a) demonstrates} that the fidelity $F$ depends strongly on the amplitude $\Omega_0$ of STIRAP and f-STIRAP, where the fidelity can be defined as $F= |\langle \Psi_0 (t_f) | \Psi (t_f) \rangle|^2$, with $|\Psi_0 (t_f) \rangle$ being the target state {(the dark state at $t=t_f$)} and $ |\Psi (t_f) \rangle$ being the final solution of time-dependent Schr\"{o}dinger equation. It is clear that the adiabatic passages do not work for small {values} of $\Omega_{0}$, due to the breakdown of adiabatic {condition}. For instance, when $\Omega_0=10$ {is chosen,} {which is much less $\Omega_{0}=100$,} as used in Fig. \ref{fig2}, the fidelity of nonlinear STIRAP is far away from {being unity}. Remarkably, the designed FF driving field {accelerates} the adiabatic passage with perfect fidelity, $F \simeq 1$, for arbitrary value of $\Omega_{0}$. In addition, as shown in Fig. \ref{fig5},  $\Omega_{0}=20$ is required for nonlinear f-STIRAP to meet the adiabatic criteria. With the assisted FF driving , the perfect population transfer
can be achieved when $\Omega_{0}=5$. 
However, one has to keep in mind that the energetic cost of STA, that is, the physical constraint on the FF driving field sets the limitation to shorten the time, relevant to quantum speed limits \cite{energycost}. We also confirm that FF scaling approach improves the stability with respect to the fluctuations {in} $T$, as shown in Fig. \ref{fig6} (b). The fidelity decreases dramatically {for the adiabatic case} when $T$ is {shortened}, {Whereas, ideally,} 
it always {remains close to unity, i.e., } $F \simeq 1$ {regardless of the value chosen for $T$,} when the FF driving field is complemented. Furthermore, the stability with respect to $\tau$, affecting the sequence of Gaussian pulses, is improved by the FF driving field as well. For instance, the fidelities are decreased down to $F = 0.9790$ and $0.9953$, respectively, for original nonlinear STIRAP and f-STIRAP, when $\tau=0.5 T$. However, with the assisted FF field, the fidelities keep $F \simeq 1$ in both protocols, where $T$ is rescaled by magnification factor $R(t)=2t$ with the assisted FF field. Moreover, we check that 
the other protocols for previously mentioned $R(t)$ keep the same feature of robustness against variation of $\Omega_{0}$ and $T$. Actually, the improvement of robustness makes sense  {as} the area of whole pump and Stokes Gaussian pulses is increased by $\pi$ (for nonlinear STIRAP) and $\pi/2$ (for nonlinear f-STIRAP) {which is} induced from the FF driving field. Interestingly, we may simplify the STA recipe in this case, by using {a} one-photon 1-3 pulse, instead of original two-photon transition \cite{chenprl105}. However, the resonant pulses, i.e. $\pi$ (or $\pi/2$) pulses, are sensitive to parameter fluctuation. {As far as physical implementations are concerned}, the transition induced from the designed FF driving field can be physically implemented by a magnetic dipole transition, if electric dipole is forbidden. Therefore, the intensity of magnetic field, {that directly couples} {the states} $|1\rangle$ and $|3\rangle$, limits the ability to shorten the time {infinitely}. By combining with the laser fields, the FF driving
field, connecting levels $|1\rangle$ and $|3\rangle$,  should be on resonance with the Raman transition,
which could be problematic due to the phase mismatch and {can easily be} avoided by shaping the pulses through unitary transformations \cite{DuNc}.

\section{Conclusion}
\label{V}
In summary, we have worked out the FF scaling of coherent control for atom and molecular BECs, with the second-order nonlinearity involved. By using the dark state in nonlinear  $\Lambda$-type system, we derive the FF driving field {which, when combined} with the modified pump and Stokes fields, {can produce} high-fidelity state conversion from an atomic BEC to a molecular one {beyond the adiabatic regime.}  Moreover, the result can be directly generalized to nonlinear  f-SITRAP for the coherent superposition of an atomic and a molecular BECs. The FF assisted STIRAP and f-STIRAP have a higher tolerance to the {fluctuations in various} parameters {such as,} the intensity and {the} width of Gaussian pulses. Besides, the original adiabatic passages are speeded up, but without populating the intermediate excited state, which prevents from the losses due to inevitable dissipation \cite{Mackie} or dephasing effect \cite{Xuexi}. 

{We must} emphasize that {the} FF scaling approach in nonlinear system are different from CD driving and IE method of STA. In {a} nonlinear system, {the eigenstates are generally} non-orthogonal {and} degenerate, which hinders the calculation of CD driving, even though the {expression} of FF field is similar. With extra parameters in the phase of dark state (\ref{darkstate}), the FF driving provides more flexibility for the atom-molecular conversion, with large area of pulses. We realize that the intermediate state $|2\rangle$ is populated in an alternative IE method \cite{Dorier}, in which one of dynamical modes of Lewis-Riesenfeld invariant for its linear counterpart is applied \cite{Benseny}. However, the price paid in the FF scaling approach is the supplement of the auxiliary coupling between states $|1\rangle$ and $|3\rangle$. {And the availability of such coupling will 
set the limitation to shorten the time.} Therefore, when it comes to experimental realization, one can pick up the suitable recipes or protocols as {discussed} above, taking the physical feasibility and limitations {into account}. Moreover, there are many choices of magnification factor $R(t)$ for the rescaled time,
which gives different amplitudes of pulses. One can 
further investigate elsewhere to optimize it with respect to the amplitude of pulses and the robustness  against parameter variations, e.g. by using analytical enhanced STA \cite{prr_eSTA} or combining other numerical recipes.  

Last but not least, the FF scaling approach is interestingly extended to {study} fast and robust control of unstable nonlinear systems, {such as}, BEC in optical lattices with the third-order Kerr-type nonlinearity \cite{LZthree}, 
other applications of coupled waveguides \cite{waveguide,Spatial-review} and frequency conversion in nonlinear optics \cite{laser} in analogous fashion.

\begin{acknowledgements}
We thank Koushik Paul for his valuable discussions and comments.  We acknowledge the support from NSFC (Grant No.
12075145), STCSM (Grants No. 2019SHZDZX01-ZX04,
No. 18010500400, and No. 18ZR1415500), Program for
Eastern Scholar, Spanish Government Grant No. PGC2018-095113-B-I00 (MCIU/AEI/FEDER, UE), and Basque
Government Grant No. IT986-16.
X.C. acknowledges the Ramón y Cajal program (Grant No. RYC2017-22482).

\end{acknowledgements}

\end{document}